\documentclass[twocolumn,apj]{openjournal}
\usepackage[T1]{fontenc}
\usepackage{microtype}
\usepackage{newtxtext,newtxmath}
\usepackage[labelsep=space]{caption,subcaption}
\usepackage[breaklinks,colorlinks,citecolor=blue,urlcolor=blue]{hyperref}
\usepackage{orcidlink}

\def\TPI{Institut f\"ur Theoretische Physik I, Ruhr-Universit\"at Bochum, 44801 Bochum, Germany}
\def\TPIV{Institut f\"ur Theoretische Physik IV, Ruhr-Universit\"at Bochum, 44801 Bochum, Germany}

\def\vb#1{\boldsymbol{#1}}
\def\dd#1{\mathop{}\!\mathrm{d}{#1}}
\def\bcdot{\boldsymbol{\cdot}}
\def\btimes{\boldsymbol{\times}}
\def\bnabla{\boldsymbol{\nabla}}
\def\Levy{\mathrm{L\acute{e}vy}}

\begin{document}

\shorttitle{Anisotropic Cosmic Ray Transport in strong MHD Turbulence}
\shortauthors{J. Lübke et al.}

\title{Anisotropic Cosmic Ray Transport in strong MHD Turbulence due to Magnetic Mirroring and Resonant Curvature Scattering}
\author{Jeremiah Lübke\,\orcidlink{0000-0001-6338-9728}$^{1}$}
\author{Frederic Effenberger\,\orcidlink{0000-0002-7388-6581}$^{1,2}$}
\author{Mike Wilbert\,\orcidlink{0000-0003-2831-1583}$^{1}$}
\author{Horst Fichtner\,\orcidlink{0000-0002-9151-5127}$^{2}$}
\author{Rainer Grauer\,\orcidlink{0000-0003-0622-071X}$^{1}$}
\affiliation{$^{1}$\TPI}
\affiliation{$^{2}$\TPIV}
\thanks{E-mail: \href{mailto:jeremiah.luebke@rub.de}{jeremiah.luebke@rub.de}}

\begin{abstract}
The transport of cosmic rays through turbulent astrophysical plasmas still constitutes an open problem.
Building on recent progress, we study the combined effect of magnetic mirroring and resonant curvature scattering on parallel and perpendicular transport.
We conduct test-particle simulations in snapshots of an anisotropic magnetohydrodynamics simulation with $\delta B/B_0\sim 1$ and record magnetic moment variation and field line curvature around pitch-angle reversals.
We find for strongly magnetized particles that
(i) pitch-angle reversals may occur either in coherent regions of the field with small variation of the magnetic moment via magnetic mirroring or in chaotic regions of the field with strong variation of the magnetic moment via resonant curvature scattering;
(ii) parallel transport can be modeled as a Lévy walk with a truncated power-law distribution based on pitch-angle reversal times;
and (iii) perpendicular transport is enhanced by resonant curvature scattering in synergy with chaotic field line separation and diminished by magnetic mirroring due to confinement in locally ordered field line bundles.
While magnetic mirroring constitutes the bulk of reversal events, resonant curvature scattering additionally acts on trajectories that fall in the loss cones of typical mirroring structures and thus provides the cut-off for the reversal time distribution.
Our results, which highlight the role of the magnetic field line geometry in cosmic-ray transport processes, are consistent with energy-independent diffusion coefficients.
We conclude by considering how energy-dependent observations could arise from an intermittently inhomogeneous interstellar medium.
\end{abstract}

\keywords{Magnetohydrodynamics -- turbulence -- ISM: cosmic rays -- diffusion}

\section{Introduction}\label{sec:introduction}

The transport of cosmic rays (CRs) is governed by turbulent magnetic fields, which are characteristic for a wide range of astrophysical plasmas \citep{Amato2018,Engelbrecht2022,Ruszkowski2023}.
The guiding paradigm since \cite{Jokipii1966} and \cite{Kulsrud1969} views turbulent fluctuations~$\delta B$ as an ensemble of waves with a given energy spectrum along a strong guide field~$B_0\gg\delta B$, where particles scatter in pitch angle due to gyro resonance. This process is characterized by a single mean free path, which depends on the parameters of the turbulence spectrum and the particle energy \citep{Mertsch2020,Reichherzer2022b}.
However, in the case of strong fluctuations~$\delta B\gtrsim B_0$, non-linear interactions are expected to facilitate a critically-balanced anisotropic cascade \citep{Goldreich1995}, which is believed to provide insufficient scattering along the parallel direction for the expected mean free paths \citep{Chandran2000a}.
The fast-mode cascade \citep{Yan2004,Yan2008,Fornieri2021} and scattering by self-excited waves, e.g., via streaming instabilities \citep{Kulsrud1969,Bell2013,Zweibel2013,Sampson2023}, have been discussed as remedies for this situation, however both of these approaches appear insufficient to fully explain the available CR data in the GeV-TeV range \citep{Kempski2022,Hopkins2022a}.
Thus, new scattering mechanisms are required to understand CR transport in strong magnetic turbulence.

One possible mechanism is given by \emph{magnetic mirroring}, where particles, as they travel along a magnetic gradient, may reverse their direction due to adiabatic conservation of their magnetic moment \citep{Noerdlinger1968,Chandran2000b,Albright2001}.
Suitable configurations of the magnetic field are known to spontaneously arise in turbulence simulations, resulting in a bouncing motion of test particles \citep{Beresnyak2011,Xu2013}.
Parallel transport due to this process was studied analytically by \cite{Lazarian2021}, who derived a corresponding pitch-angle diffusion coefficient in the fast magneto-sonic case, and numerically by, e.g., \cite{Zhang2023a}, \cite{Barreto-Mota2025} and \cite{Xiao2025}.
Further, in combination with chaotic separation of field lines, this bouncing motion is believed to contribute to perpendicular transport across field lines \citep{Lazarian2021,Hu2022}.
This is supplemented by studies of CR behavior in magnetic traps \citep{Tharakkal2023,Lopez-Barquero2025}, as well as  by a Vlasov-Fokker-Planck description by \cite{Bell2025}, which combines mirroring and gyro-resonant scattering.

Another recently proposed mechanism, termed \emph{resonant curvature scattering}, states that particles cannot follow along field lines that exhibit sharp bends on scales similar to or smaller than their gyro radii and thus experience sudden changes of their magnetic momentum \citep{Lemoine2023,Kempski2023}.
The distributions of field line curvature exhibit power-law tails \citep{Yang2019,Bandyopadhyay2020}, so instances of high curvature have a small volume-filling fraction, leading to a broad distribution of free path lengths of CRs between scattering events \citep{Liang2025,Lubke2025b,Kempski2025b}. 
Pitch-angle scattering in curved field configurations has also previously been studied by, e.g., \cite{Gray1982}, \cite{Young2008} and \cite{Malara2023}.

Since magnetic mirroring and heavy tails of the field line curvature distribution are observed for a wide range of plasma setups, CR transport should be affected by both mechanisms.
In this paper, we study how magnetic mirroring and resonant curvature scattering govern the parallel and perpendicular transport of highly magnetized, i.e., low-energy, test particles in strong Alfvénic turbulence.
Specifically, particles interacting via these two mechanisms with the magnetic field may experience pitch-angle reversals, where the distribution of times between consecutive reversals exhibits a truncated power-law shape, from which the parallel diffusion coefficient can be computed.
Further, reversals due to mirroring tend to diminish cross-field transport, while reversals due to resonant curvature scattering tend to enhance it, leading to a modified perpendicular diffusion coefficient, which is to first order given by field line wandering.

The paper is organized as follows: Section~\ref{sec:simulation} introduces the numerical setup, Section~\ref{sec:reversals} discusses the scattering mechanisms for pitch-angle reversals, Section~\ref{sec:Levy} introduces a Lévy-walk model for parallel CR transport and Section~\ref{sec:perp} presents statistics of perpendicular transport arising from reversal events.
We contextualize our results in Section~\ref{sec:energy-independent-transport}, where we discuss the possibility of energy-independent parallel diffusion coefficients, and in Section~\ref{sec:inhomogeneous-ISM}, where we discuss how energy-dependent observations could arise from an inhomogeneous interstellar medium (ISM).
We then conclude our paper in Section~\ref{sec:conclusion}.

\section{Test-Particle Simulation}\label{sec:simulation}

\begin{figure*}
    \begin{subfigure}{.49\linewidth}
        \includegraphics[width=\linewidth]{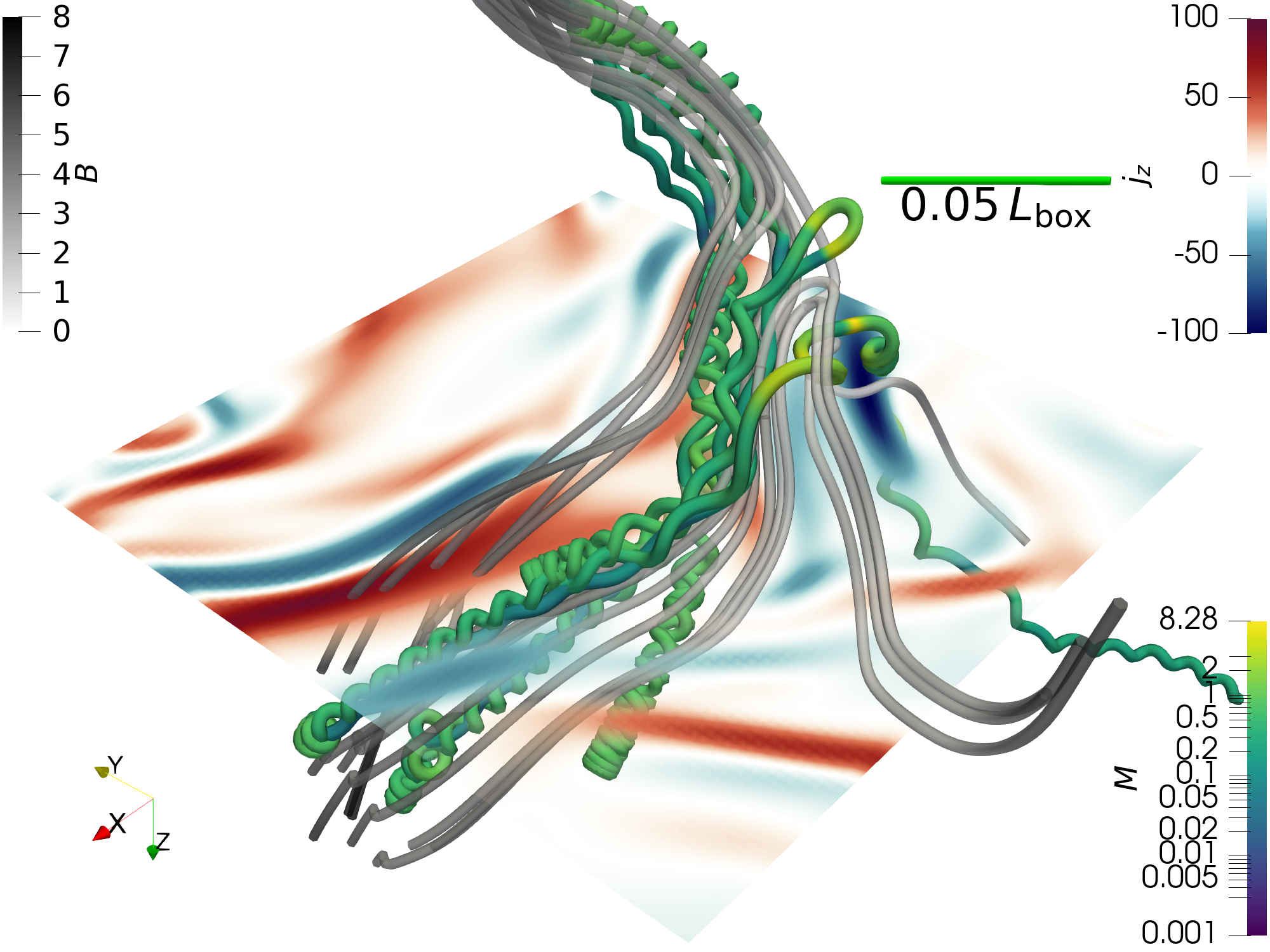}
        \subcaption{}
    \end{subfigure}\hfill%
    \begin{subfigure}{.49\linewidth}
        \includegraphics[width=\linewidth]{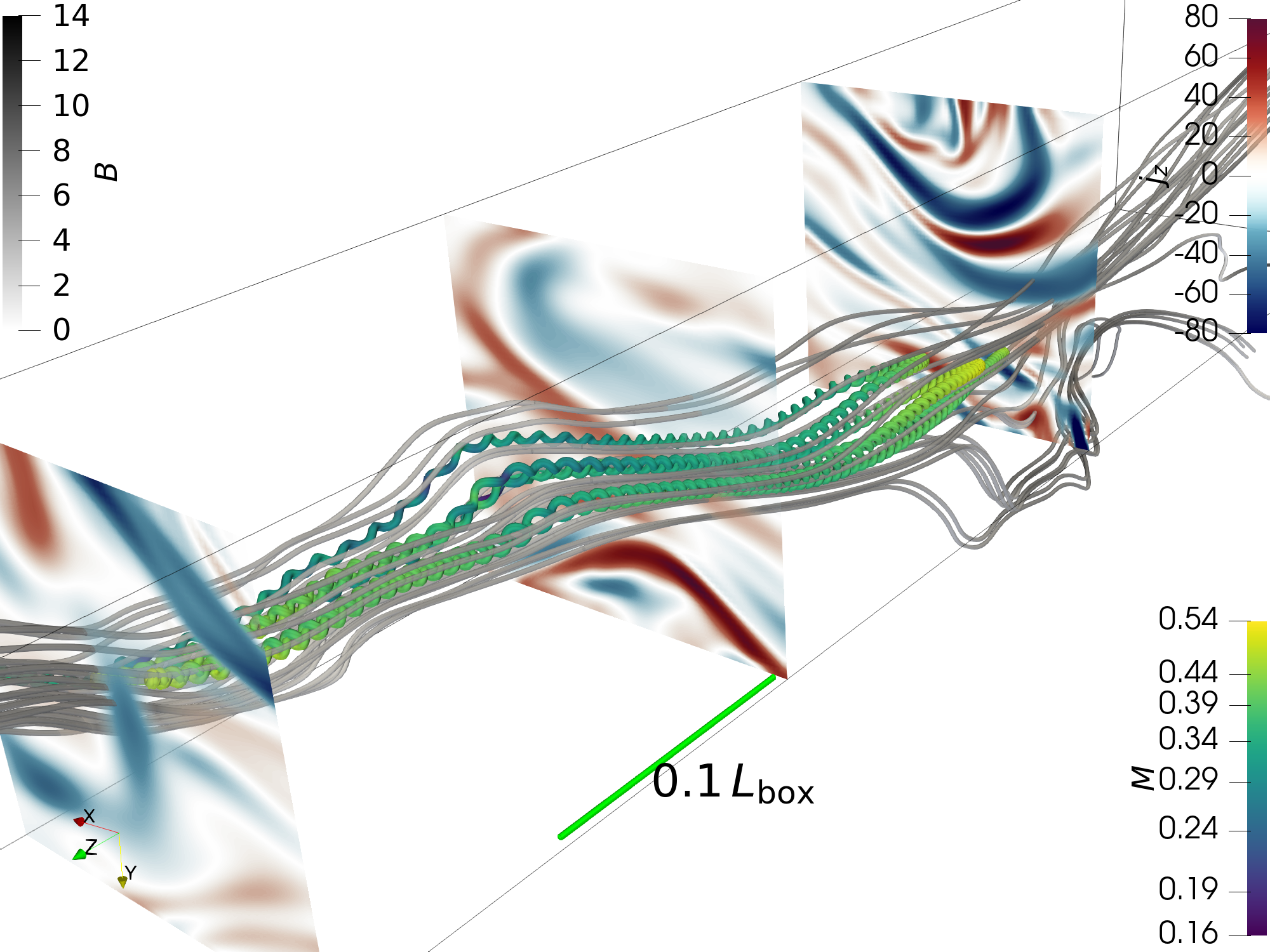}
        \subcaption{}
    \end{subfigure}
    \caption{Example test-particle trajectories illustrating the two prevalent scattering mechanisms in our simulation: (a)~\emph{resonant curvature scattering} characterized by strong variations of the magnetic moment~$M$ upon encountering sharply bent field lines, and (b)~\emph{magnetic mirroring} characterized by pitch-angle reversals with small variations of~$M$ induced by particles traveling along field line configurations with slowly increasing magnetic field strengths.
    The structure of the magnetic field is illustrated by field line trajectories and slices of the out-of-plane current density~$j_z$.
    Interactive versions of these figures are available online via sketchfab: (a) \url{https://skfb.ly/pDwuD}, (b) \url{https://skfb.ly/pDwGD}.}
    \label{fig:trajectories}
\end{figure*}

We consider a simulation of forced incompressible visco-resistive magnetohydrodynamic (MHD) turbulence. The magnetic field~$\vb{B}=\vb{B}_0+\delta\vb{B}$ consists of a constant background component~$\vb{B}_0=B_0\hat{\vb{z}}$ and a fluctuating component~$\delta\vb{B}$ with~$\delta B_\mathrm{rms}/B_0\approx 1$.
This choice is motivated by the Alfvén Mach number~$\mathcal{M}_A=\frac{u_\mathrm{rms}}{\sqrt{\smash[b]{B_0^2+\delta B_\mathrm{rms}^2}}}\approx 0.7$ (using Alfvénic units~$[B]=[u]$ and energy equipartition~$\delta B_\mathrm{rms}\approx u_\mathrm{rms}$), which appears reasonable for the ISM \citep{Crutcher1999,PlanckCollaboration2016,Ntormousi2024}.
The simulation box measures~$L_x\times L_x\times L_z$ on a~$1024^3$ grid, with perpendicular length~$L_x$ and parallel length~$L_z=\sqrt{2}\,L_x$.
See Appendix~\ref{app:mhd} for technical details.

We trace test particles, parametrized by the normalized root mean squared (RMS) gyro frequency~$\hat{\omega}_g$, through eight statistically saturated and independent snapshots of this simulation by integrating the Newton-Lorentz equations
\begin{equation}
    \dot{\vb{X}}_t=\vb{V}_t, \quad \dot{\vb{V}}_t=\hat{\omega}_g\,\vb{V}_t\btimes\vb{B}(\vb{X}_t),
    \label{eq:lorentz}
\end{equation}
with the volume-preserving Boris scheme \citep{boris:1971}. We employ a fixed time step~$\Delta t=0.01\times2\pi\,\hat{\omega}_g^{-1}$ and evaluate the magnetic field at the particle position by trilinear interpolation.
Field line trajectories are obtained by integrating the ordinary differential equation, parametrized by the arc length~$s$,
\begin{equation}
    \dot{\vb{X}}_s=\hat{\vb{B}}(\vb{X}_s),
    \label{eq:fieldlines}
\end{equation}
where~$\hat{\vb{B}}=\vb{B}/B$ denotes the unit vector of the magnetic field.

The test-particle approach is justified by assuming particle speeds~$V$ much higher than the RMS plasma speed~$u_\mathrm{rms}$, which implies that the electric field is negligible and particle energies are conserved.
The RMS gyro frequency is given by
\begin{equation}
    \hat{\omega}_g=\frac{qB_\mathrm{rms}L_x}{\gamma mc},
    \label{eq:omega}
\end{equation}
which encodes the typical gyro period~$T_g/(L_x/c)=2\pi\,\hat{\omega}_g^{-1}$, normalized gyro radius~$\hat{r}_g=r_g/L_x=\frac{\pi}{2}\,\hat{\omega}_g^{-1}$ and
particle energy~$E=\gamma mc^2=qB_\mathrm{rms}L_xc/\hat{\omega}_g$.
Our code normalizes the system to~$c=1$,~$B_\mathrm{rms}=1$ and~$L_x=1$.

We consider eight logarithmically spaced values for~$\hat{\omega}_g\in\{512,\cdots,4\}$, corresponding to normalized gyro radii~$\hat{r}_g\in\{0.003,\cdots,0.393\}$, and simulate~$40\,000$ trajectories for~$100\,000$ gyro periods in each snapshot and for each~$\hat{\omega}_g$.
Further, we trace~$400\,000$ field lines up to arc length~$s=100$ in each snapshot.

\section{Transport Behavior}
\subsection{Magnetic Mirroring and Resonant Curvature Scattering}\label{sec:reversals}
Particles are magnetized, i.e., tied to magnetic field lines, when their gyro radius~$r_g$ is smaller than the scale over which the magnetic field varies.
While turbulence is generally characterized by a wide range of scales, the field line curvature radius~$\kappa^{-1}$ provides a local measure of the magnetic field variation \citep{Lemoine2023,Kempski2023,Lubke2025b}, so we consider particles as locally magnetized if~$r_g\kappa<1$ and denote their degree of magnetization by the probability~$p(r_g\kappa<1)$.
Strongly magnetized particles are further characterized by a conserved magnetic moment~$M=(1-\mu^2)/2B$, where~$\mu=\hat{\vb{V}}\bcdot\hat{\vb{B}}$ denotes the pitch-angle cosine.
Their motion along magnetic field lines may be interrupted by pitch-angle reversals due to magnetic mirroring, which conserves~$M$, or by intermittent encounters with sharply bent field lines, where they become locally unmagnetized and experience strong variations in~$M$.
Figure~\ref{fig:trajectories} shows examples of such events.

\begin{figure}
    \centering
    \includegraphics[width=\linewidth]{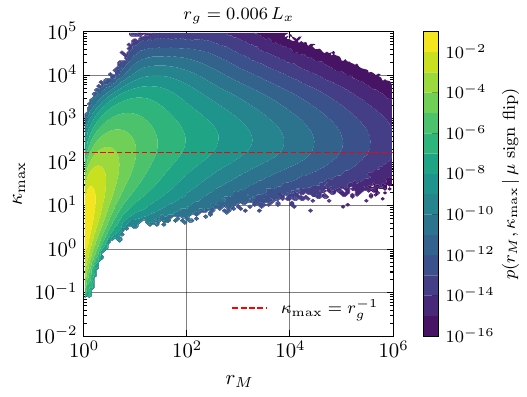}
    \caption{Joint probability distribution~$p(r_M,\kappa_\mathrm{max})$ of magnetic moment variation~$r_M=M_\mathrm{max}/M_\mathrm{min}$ and maximum field line curvature~$\kappa_\mathrm{max}$ experienced by test particles with~$r_g=0.006\,L_x$ during~$2\,T_g$-intervals centered around pitch-angle reversal events.
    The red line indicates the resonant curvature condition, above which unmagnetized scattering is expected to dominate.}
    \label{fig:mu-reversals}
\end{figure}

To understand the nature of these perturbations, we record the magnetic moment variation~$r_M=M_\mathrm{max}/M_\mathrm{min}$ and the maximum field line curvature~$\kappa_\mathrm{max}$ experienced by particles during~$2\,T_g$-intervals centered around pitch-angle reversal events, i.e., changes of the pitch angle through~$90^\circ$.
Figure~\ref{fig:mu-reversals} depicts the joint probability distribution~$p(r_M,\kappa_\mathrm{max})$ for highly magnetized low-energy particles with~$r_g=0.006\,L_x$ and~$p(r_g\kappa<1)\approx 0.98$.
The core of the distribution, which is centered around~$r_M\approx1$ and~$\kappa_\mathrm{max}\approx3$, consists of slow magnetized pitch-angle reversals caused by magnetic mirroring. These events require sufficiently coherent field line bundles so that particles do not become unmagnetized.
However, when particles do encounter tangled field line bundles with sharp curvatures, they become unmagnetized and experience strong variations of~$M$ on short time scales. These events constitute resonant curvature scattering and populate the extended tail of the distribution towards high~$r_M$ and~$\kappa_\mathrm{max}$.

We recall that field line curvature and magnetic field strength are anti-correlated as~$B\sim\kappa^{-1/2}$ \citep{Schekochihin2004a,Kriel2025}. 
Field line bundles with strong~$B$ tend to be coherent with low~$\kappa$, where magnetic mirroring may occur, while field line bundles with weak~$B$ tend to be highly tangled with high~$\kappa$, where resonant curvature scattering may occur.

\subsection{Parallel Transport}\label{sec:Levy}
\begin{figure*}
    \begin{subfigure}{.5\linewidth}
        \includegraphics[width=\linewidth]{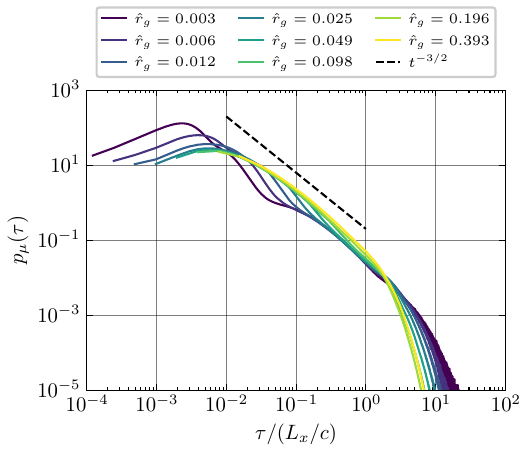}
        \subcaption{}
    \end{subfigure}\hfill%
    \begin{subfigure}{.5\linewidth}
        \includegraphics[width=\linewidth]{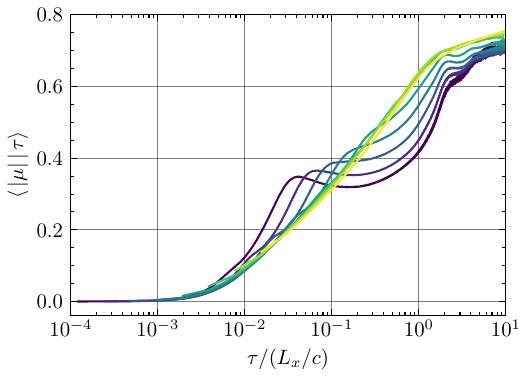}
        \subcaption{}
    \end{subfigure}
    \caption{(a) Distributions~$p_\mu(t)$ of durations~$\tau$ between pitch-angle reversals, which exhibit a power-law scaling similar to the first-passage scaling~$t^{-3/2}$ (indicated for reference) over about two orders of magnitude, as well as energy-dependent cut-off scales.
    (b) Average of the absolute value of the pitch-angle cosine~$\mu$ conditional on segment duration~$\tau$. Particles with large~$\mu$ have a higher probability of falling into the loss cone of traversed structures and thus avoid mirroring, which leads to larger~$\langle|\mu|\,|\,\tau\rangle$ at longer time scales.}
    \label{fig:pt}
\end{figure*}

We now examine how well the statistics of pitch-angle reversals describe the parallel transport of our test particles.
For this purpose, we assume a Lévy walk where particles travel along field lines with a constant speed~$v_\Levy$ until a reversal event occurs, from where they continue traveling in the opposite direction along their field line until another reversal event occurs, and so on.
The durations~$\tau$ of these segments are independently sampled from the probability distribution~$p_\mu(\tau)$ of durations between pitch-angle reversals in our test-particle data.
Provided that the first two moments of~$p_\mu(\tau)$ exist, the diffusion coefficient of this Lévy walk is given by \citep{Zaburdaev2015}
\begin{equation}
    D^\infty_\Levy=\frac{\langle x^2\rangle}{2\langle \tau\rangle}=\frac{\langle v^2_\Levy\,\tau^2\rangle}{2\langle \tau\rangle},
    \label{eq:DLevy}
\end{equation}
where the displacement~$x$ is given by construction as~$x=v_\Levy\,\tau$.

\begin{figure}
    \centering
    \includegraphics[width=\linewidth]{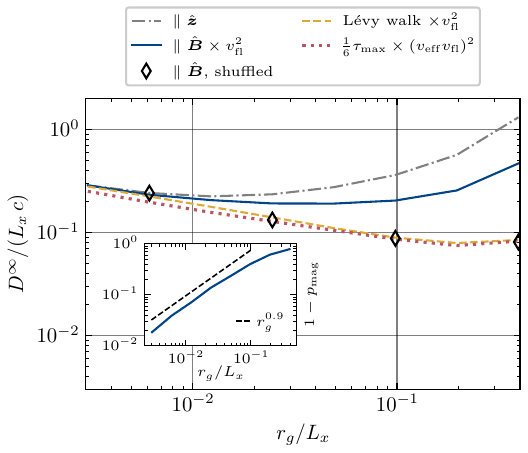}
    \caption{Diffusion coefficients as functions of the normalized gyro radius, parallel to the global direction of the background field~$\hat{\vb{z}}$ and parallel to the local direction of the magnetic field~$\hat{\vb{B}}$,
    as well as exact and approximate diffusion coefficients computed from the Lévy-walk model. The exact value takes the entire reversal time distribution~$p_\mu(\tau)$ into account, while the approximate value is given by the cut-off time scale~$\tau_\mathrm{max}$ and the effective Lévy speed~$v_\mathrm{eff}$.
    Finally, the diffusion coefficients of the shuffled~$\mu_t$--time-series~($\lozenge$) aid to resolve the discrepancy between~$D^\infty_{\parallel\hat{\vb{B}}}$ and~$D^\infty_\Levy$.
    Quantities in the frame of the local field direction~$\hat{\vb{B}}$ are scaled by the field line speed~$v_\mathrm{fl}$.
    The inset shows one minus the degree of magnetization~$1-p_\mathrm{mag}=1-p(r_g\,\kappa<1)$.}
    \label{fig:mfp}
\end{figure}

Notably, the probability distributions~$p_\mu(\tau)$, which are plotted in Figure~\ref{fig:pt}a, reveal a power-law behavior over about two orders of magnitude, which scales roughly as~$\sim\tau^{-3/2}$, and a cut-off time scale~$\tau_\mathrm{max}$, which decreases with increasing particle energy.
Particles exhibit an extended ballistic phase on the time scales of the power-law scaling and become diffusive beyond~$\tau_\mathrm{max}$.
The power-law scaling is reminiscent of the classical first-passage scaling of a Markovian walker on a one-dimensional interval \citep{Liang2025}.
In our previous work on isotropic turbulence \citep{Lubke2025b}, we observed similar scaling and cut-off behavior for durations of magnetized motion.
Similar power-law distributions for reversal times have also been observed in intermittent synthetic turbulence by \citet{Pucci2016}.

We can estimate Equation~(\ref{eq:DLevy}) with a constant effective Lévy speed~$v_\mathrm{eff}$ and a simple truncated power-law distribution~$p(\tau)\sim\tau^{-\alpha}$ for~$\tau\in(\tau_\mathrm{min},\tau_\mathrm{max})$, $\tau_\mathrm{min}\ll\tau_\mathrm{max}$ and~$\alpha\in(1,2)$ as
\begin{equation}
    D^\infty_\Levy\approx\frac{1}{2}v^2_\mathrm{eff}\frac{2-\alpha}{3-\alpha}\tau_\mathrm{max}\overset{\alpha=\frac{3}{2}}{=}\frac{v^2_\mathrm{eff}\tau_\mathrm{max}}{6},
    \label{eq:DLevyApprox}
\end{equation}
which reveals that the diffusion coefficient is given by the largest involved time scales \citep{Kempski2025b}.
To calculate Equation~(\ref{eq:DLevy}) exactly, we have to consider the Lévy speed with dependence on the segment length~$\tau$ as~$v_\Levy(\tau)=c\,\langle|\mu|\,|\,\tau\rangle$, due to magnetic mirroring.
Specifically, for a magnetic structure with field strength~$B_\mathrm{min}$ at its center and~$B_\mathrm{max}$ at its end, particles will undergo magnetic mirroring if their pitch-angle cosine at the center of the structure fulfills~$|\mu|\leq\mu_\mathrm{mirror}=\sqrt{1-B_\mathrm{min}/B_\mathrm{max}}$ \citep{Chen2016_with_publisher}.
Assuming a uniform distribution for~$\mu_\mathrm{mirror}$, particles with small~$\mu$ have a higher probability for a reversal event and consequently the tails of~$p_\mu(\tau)$ are comprised of particles with larger~$\mu$, which are more likely to fall into the loss cone of the mirroring structures.
The numerically measured data for~$\langle|\mu|\,|\,\tau\rangle$ is plotted in Figure~\ref{fig:pt}b, which confirms our consideration.

As shown in Figure~\ref{fig:mfp}, the approximate values given by Equation~(\ref{eq:DLevyApprox}) reproduce the energy-dependency of the exact values given by Equation~(\ref{eq:DLevy}) rather well,
where we estimate~$v_\mathrm{eff}$ from the ratio of Equations~(\ref{eq:DLevy}) and~(\ref{eq:DLevyApprox}) as~$v_\mathrm{eff}\approx0.65$ for all considered energies (for reference,~$v_\mathrm{eff}=0.5$ would be expected for an isotropic pitch-angle distribution and in the absence of the loss-cone selection effect).
This shows that the exact diffusion coefficients, despite non-trivial~$\langle|\mu|\,|\,\tau\rangle$, are primarily determined by the largest involved time scales (up to a factor~$v_\mathrm{eff}^2$).

For the test particle data, we compute diffusion coefficients from mean squared displacements (MSDs) as~$D^\infty=\lim_{t\to\infty}\langle\Delta X^2(t)\rangle/2t$, once for the usual global parallel case~$\Delta X_{\parallel\hat{\vb{z}}}(t)=X_{z,t}-X_{z,0}$ and once for the local field-aligned case~$\Delta X_{\parallel\hat{\vb{B}}}(t)=\int_0^t\vb{V}_{t'}\bcdot\hat{\vb{B}}_{t'}\dd{t'}=c\int_0^t\mu_{t'}\dd{t'}$, which corresponds to the pitch-angle dynamics modeled by the Lévy walk.
Displacements in the field-aligned frame of reference and the global frame of reference relate to each other as~$\Delta X_{\parallel\hat{\vb{z}}}(t)\approx v_\mathrm{fl}\,\Delta X_{\parallel\hat{\vb{B}}}(t)$, where we take the parallel field line displacement~$\langle\Delta X_{\parallel,\mathrm{fl}}^2(s)\rangle=v_\mathrm{fl}^2s^2$ into account, which is ballistic and characterized by the field line speed~$v_\mathrm{fl}\approx 0.7$.

The comparison of test-particle and Lévy-walk diffusion coefficients in Figure~\ref{fig:mfp} shows that the Lévy walk provides a good description for highly magnetized low-energy particles, however it also reveals growing discrepancy for increasing particle energies.
This discrepancy is puzzling at first given that~$\tau_\mathrm{max}$ decreases, but can be explained by decreasing magnetization, as shown in the inset of Figure~\ref{fig:mfp}. Such particles are only weakly coupled to magnetic field lines and may thus travel over tangled field line bundles without their motion being significantly affected.
In other words, the higher momentum of less magnetized particles introduces strong correlations between successive pitch-angle reversals, which lead to larger diffusion coefficients despite decreasing cut-off time scales~$\tau_\mathrm{max}$.
We test this explanation explicitly by dividing time series of the pitch-angle cosine~$\mu_t=(\mu_{t_0},\cdots,\mu_{t_n})$ from our test-particle simulation into connected segments with the same sign and randomly re-ordering these segments.
This removes correlations between successive pitch-angle reversals and leads to agreement of the resulting diffusion coefficient~$D^\infty_{\parallel\hat{\vb{B}},\mathrm{shuffled}}$ with the Lévy-walk model.

\subsection{Perpendicular Transport}\label{sec:perp}

\begin{figure}
    \centering
    \includegraphics[width=\linewidth]{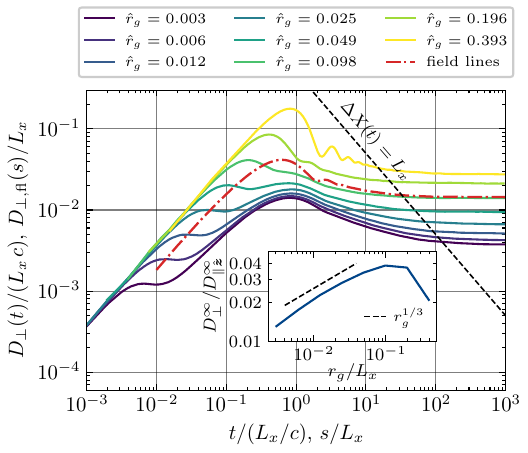}
    \caption{Running perpendicular diffusion coefficients of test particles and field lines. As indicated, convergence to diffusion occurs for displacements beyond the perpendicular box size. The inset relates the resulting perpendicular diffusion coefficients to the parallel diffusion coefficients.}
    \label{fig:rdc-perp}
\end{figure}

The background field imposes a preferred direction of the field vectors with~$\langle B_z\rangle=B_0$, leading to field line transport, which is ballistic in the parallel direction and suppressed in the perpendicular direction.
Charged particles, preferably following along field lines, are also affected by this anisotropy, as illustrated by the ratios of parallel and perpendicular diffusion coefficients shown in the inset of Figure~\ref{fig:rdc-perp}.
The main part of the figure shows perpendicular running diffusion coefficients~$D_{\perp}(t)=\langle(X_{x,t}-X_{x,0})^2+(X_{y,t}-X_{y,0})^2\rangle/2t$ of test particles and field lines (parametrized with the arc length~$s$ instead of time~$t$), which reveal transient subdiffusion~$D_\perp(t)\sim t^{\alpha-1}$ with~$\alpha<1$ on time scales~$t\sim(1\cdots100)\,L_x/c$, before converging to normal diffusion with~$\alpha=1$ after crossing the size of the simulation box.
This subdiffusive behavior is present for field lines and particles with medium to low energies, while for high-energy particles effects of the large gyro radii dominate.
Field lines are transiently subdiffusive because their perpendicular displacement is inhibited by the imposed background field and due to slow remaining perpendicular wandering, they may decorrelate from their initial condition and thus recover normal diffusion on long arc length scales~$s\gtrsim 10^2\,L_x$.
Sufficiently magnetized particles are affected by field line wandering, so they resemble the subdiffusive field line behavior. This is distinct from compound subdiffusion observed for CR transport in isotropic turbulence of the fluctuating dynamo \citep{Lubke2025b}.

\begin{figure}
    \centering
    \includegraphics[width=\linewidth]{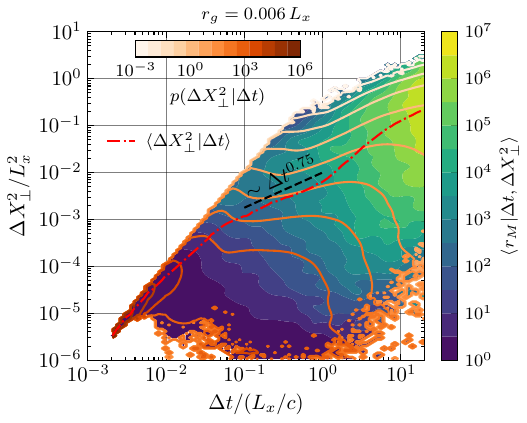}
    \caption{Test particle excursions, defined as trajectories that return to their initial~$z$-coordinate, characterized by their total magnetic moment variation~$r_M$, duration~$\Delta t$ and final squared perpendicular displacement~$\Delta X^2_\perp$.
    Also indicated are the distribution and average of~$\Delta X^2_\perp$ conditional on~$\Delta t$.}
    \label{fig:conditional}
\end{figure}

However, field line wandering gives only a first-order approximation of perpendicular particle transport.
For instance, the subdiffusive behavior becomes more pronounced for smaller particle energies.
Such deviations may be caused by confinement \citep{Chandran2000b} or cross-field transport \citep{Desiati2014}.
To understand the role of pitch-angle reversals in confinement and cross-field transport, we collect particle trajectories that return to their initial~$z$-coordinate \citep[compare also the approach in][]{Laitinen2017b}.
We characterize these excursions, which may contain multiple pitch-angle reversals, by their duration~$\Delta t=t_\mathrm{return}-t_0$, squared perpendicular displacement~$\Delta X_\perp^2=(X_{x,t_\mathrm{return}}-X_{x,0})^2+(X_{y,t_\mathrm{return}}-X_{y,0})^2$ and total magnetic moment variation~$r_M=M_\mathrm{max}/M_\mathrm{min}$.
Building on Section~\ref{sec:reversals}, we consider~$r_M$ as an indicator for the scattering mechanism that mostly affects an individual excursion, i.e., small~$r_M$ indicates reversal by mirroring, while large~$r_M$ indicates reversal by curvature scattering.

We consider the results for highly magnetized particles with~$r_g=0.006\,L_x$ that are plotted in Figure~\ref{fig:conditional}.
Firstly,~$r_M$ increases as the duration of the excursions increases, due to longer trajectories having more opportunities for resonant curvature scattering.
Also,~$r_M$ increases as the perpendicular displacement increases, indicating that resonant curvature scattering enables cross-field transport \citep{Lemoine2023,Kempski2023}, which leads to enhanced perpendicular displacement in synergy with chaotically separating field lines \citep{Huang2014}.
On the other hand, excursions with small~$r_M$ due to magnetic mirroring exhibit very small perpendicular displacements, even after intermediate durations~$\Delta t\sim 1\,L_x/c$.
This indicates that magnetic mirroring inhibits perpendicular transport, as particles are confined in the coherent mirroring structures of the magnetic field.
Analogous to Section~\ref{sec:reversals}, regions with a weak field and large curvature likely exhibit a high degree of chaotic field line separation, which enhances cross-field transport, while regions with a strong field and small curvature are composed of ordered field line bundles, which enhances confinement by magnetic mirroring.
The overlaid probability density~$p(\Delta X_\perp^2|\Delta t)$ in Figure~\ref{fig:conditional} indicates a predominance of confinement by magnetic mirroring on intermediate time scales, which leads to the indicated subdiffusive scaling of returned particles and explains the diminished perpendicular transport for~$r_g\lesssim0.049\,L_x$ compared to the field line case shown in Figure~\ref{fig:rdc-perp}.

Our finding of subdiffusive perpendicular transport due to confinement by magnetic mirrors seems to differ from previous works \citep{Lazarian2021,Hu2022}, which claim superdiffusive perpendicular transport in conjunction with magnetic mirroring.
\cite{Zhang2023a} study these ideas numerically, including in incompressible Alfvénic turbulence similar to our setup.
This discrepancy can partly be explained by the different approaches to measuring the dispersion of particles: these authors consider the relative distance between pairs of particles in a beam~$\delta X_\perp^{(1,2)}(t)=\|\vb{X}_{\perp,t}^{(1)}-\vb{X}_{\perp,t}^{(2)}\|$, while we consider the absolute displacement of single particles~$\Delta X_{\perp}^{(1)}(t)=\|\vb{X}_{\perp,t}^{(1)}-\vb{X}_{\perp,0}^{(1)}\|$.
While both quantities provide valid ways to measure the time-dependent spread of a distribution of particles, it is not entirely clear how they relate to each other and thus this topic deserves further study.

\section{Discussion}\label{sec:discussion}
\subsection{Energy-independent Transport}\label{sec:energy-independent-transport}

We have argued in Section~\ref{sec:Levy} that parallel transport of highly magnetized particles arises primarily from pitch-angle reversals facilitated by magnetic mirroring and resonant curvature scattering.
Classical gyro-resonant scattering, on the other hand, does not appear to play a significant role in this regime.
The reason for this is two-fold: First, diffusion due to gyro-resonance relies on the accumulation of sufficiently many small-angle scattering events, which can be interrupted by the large-angle scattering events discussed above.
Especially, mirroring is expected to dominate over gyro-resonance at smaller pitch-angle cosines~$|\mu|\lesssim0.5$.

Second, quasi-linear theory assumes that turbulence consists of space-filling fluctuations~$\delta\vb{B}(\vb{k})$ with amplitudes given by the energy spectrum, which facilitate continuing gyro-resonant scattering.
However, in strong turbulence,~$\delta B\gtrsim B_0$, non-linear interactions give rise to intricate and inhomogeneous field line geometry consisting of chaotic and coherent regions accompanied by intense current sheets \citep{Greco2008,Pezzi2022,Dong2022,Zhou2020}.
Individual field line bundles that carry only a subset of the active wave numbers~$\vb{k}$ only act on particles with a narrow pitch-angle range as per the resonance condition~$k \mu r_g \sim 1$.
Those weakly scattering particles then travel along the field line bundle with free paths comparable to the length of the structure until a sharp curvature event is encountered.
Strongly aligned particles with~$|\mu|\approx 1$, that fall in the loss cones of most structures, may even traverse the simulation box multiple times before encountering a reversal event, as can be seen for~$r_g=0.006\,L_x$ with~$\tau_\mathrm{max}\approx3.95\,L_z/c$, which corresponds to a parallel displacement~$\Delta X_{\parallel,\mathrm{max}}=cv_\mathrm{fl}\tau_\mathrm{max}\approx2.79\,L_z$.

Assuming gyro resonance is inefficient, the distribution of reversal times or free paths is, on one hand, given by magnetic mirroring, which is independent of the particle energy and resembles the distribution of lengths of coherent field line bundles.
On the other hand, the free paths due to resonant curvature scattering resemble the spatial distribution of sharp curvature regions that may be highly clustered.
Specifically, we expect that the sharpest curvature events are closely surrounded by slightly less curved field lines \citep[compare the model of singular structures by][]{She1994}.
Thus, individual sharp curvature regions may act equally as scattering locations for a wide range of particle energies and thereby render resonant curvature scattering effectively energy-independent.
To better understand this issue, a study of the curvature correlation function along magnetic field lines would be highly interesting.

Our results for the parallel diffusion coefficients in the low-energy regime depend only weakly on the particle energy and, as shown by \cite{Kempski2025b}, this dependency becomes even weaker with increasing resolution, so numerical results appear at least consistent with energy-independent parallel transport.
If this scenario applies to galactic turbulence, the wandering of magnetic field lines plays a central role in understanding CR confinement times \citep{Heesen2023,Pezzi2024}.

A numerical approach, which simultaneously captures the most extreme scattering events and the free paths~$\lambda_\mathrm{max}$ between them, requires massive effort to resolve both extremely small curvature radii as well as multiple parallel correlation lengths.
Imposing, for instance,~$\kappa^{-1}/L_\parallel\lesssim 10^{-3}$ and~$L_z/L_\parallel\sim 10^{1}$ implies resolutions exceeding~$10\,000^3$ grid points, where state-of-the-art simulations \citep{Dong2022,Beattie2024,Kempski2025b} could be supplemented by instanton-based synthetic turbulence approaches \citep{Schorlepp2025a} \citep[see also][]{Lubke2024,Martin2025}.

The underlying idea of our work, that CR transport is governed by the geometry of the turbulent magnetic field, could be derived \emph{ab initio} via the Mori-Zwanzig formalism \citep{mori:1965,zwanzig:1973}, where the relevant parts of the particle motion are extracted by a projection operator, which represents the guiding role of the field line structure characterized by intense current sheets and aligned fluctuations \citep{Grauer1994,Mallet2015}.
Orthogonally, rareness of scattering events, as evident in the broad probability distribution of reversal times, may lead to subdiffusion of the pitch angle, which in turn leads to superdiffusion in space \citep{Zimbardo2020} and thus warrants a (truncated) fractional transport theory \citep{Magdziarz2007,Manke2019,Aerdker2025}.

Lastly, even for a moderately strong background field, perpendicular field line transport is already significantly suppressed, which, in combination with confinement by magnetic mirroring, leads to strongly diminished perpendicular transport of CRs.
This suggests that cross-field transport plays a subordinated role for highly magnetized particles and underlines the need for anisotropic transport models \citep{Effenberger2012a,Dorner2024,Kleimann2025}.
Statements on the energy-dependency of perpendicular transport are not reliable at the currently available numerical resolution, because deviations from field line wandering depend more delicately on the small-scale structure of the field, in contrast to parallel transport governed by the largest excursions.

\subsection{A possible mechanism for energy-dependent transport}\label{sec:inhomogeneous-ISM}
Our conjecture of energy-independent parallel transport stands in tension with available observations, such as the boron-to-carbon (B/C) ratio, which implies that diffusion coefficients of galactic CRs in the GeV-TeV range  scale as~$D\sim r_g^{1/3}$ \citep{Aguilar2016}.
To address this issue, we note at first that our diffusion coefficients were determined from a box containing about~$L_z/L_\parallel\approx 2.4$ parallel correlation lengths, while the galactic disk with a diameter of~$\sim30\,\mathrm{kpc}$ contains a large number of turbulent correlation cells, assuming~$L_\parallel\approx 100\,\mathrm{pc}$.
We thus sketch a possible scenario in the spirit of \cite{Butsky2024} \citep[see also][]{Reichherzer2025,Ewart2025}, which could produce energy-dependent observations on large scales despite energy-independent processes on small scales.

For this purpose, we postulate the expression of the energy-independent parallel diffusion coefficient
\begin{equation}
    D_0=\frac{1}{2}cv_\mathrm{fl}L_\parallel,
\end{equation}
dependent on the field line speed~$v_\mathrm{fl}$ (which encodes~$\delta B/B_0$) and parallel correlation length~$L_\parallel$.
For our MHD simulation with~$v_\mathrm{fl}\approx 0.7$ and~$L_\parallel\approx 0.59\,L_x$, we have~$D_0\approx0.21\,cL_x$, which is consistent with our numerical results (compare Figure~\ref{fig:mfp}).
We then apply this expression to a simplified ISM, which is filled with structured Alfvénic turbulence \citep{Ntormousi2024}, uniformly characterized by~$\delta B/B_0\approx1$ and parallel correlation scale~$L_\parallel=100\,\mathrm{pc}$.
Within this medium, we imagine localized turbulent patches of size~$l<L_\parallel$, with~$\delta B/B_0\approx 1$ and correlation scale~$L_{\parallel,\mathrm{patch}}\approx l$.
These patches are distributed intermittently throughout the domain with a small volume-filling fraction~$q$, and their sizes follow a power-law distribution~$p(l)\sim l^{-\alpha}$ over a range~$(l_\mathrm{min},l_\mathrm{max})$ with~$l_\mathrm{min}\ll l_\mathrm{max}\lesssim L_\parallel$.

CR protons with~$r_g<L_\parallel$ are usually transported through this medium with~$D_0$, until they encounter a patch of size~$l>r_g$, where they are locally transported with~$D_1(l)=\frac{1}{2}cv_\mathrm{fl}l$.
Due to~$q\ll 1$, small-angle scattering caused by patches with~$l<r_g$ is suppressed and the respective particles continue to be transported with~$D_0$.
Specifically, small-angle deflections caused by those patches are quickly overridden by the ambient turbulence and can thus not accumulate.
The effective diffusion coefficient for such an inhomogeneous medium is given by the harmonic mean, which we write as
\begin{align}
    D(r_g)&=\bigg[
    \underbrace{(1-q)D_0^{-1}}_{\substack{%
        \text{transport} \\ \text{without patches}}}
    +\underbrace{qD_0^{-1}\int_{l_\mathrm{min}}^{r_g}p(l)\dd{l}}_{\substack{%
    \text{small patches} \\ \text{without effect}}}
    \nonumber \\ & \qquad
    +\underbrace{q\int_{r_g}^{l_\mathrm{max}}D_1^{-1}(l)\,p(l)\dd{l}}_{\substack{%
    \text{transport in relevant patches}}}
    \bigg]^{-1} \nonumber \\
    &=\left[(1-qf(r_g))D_0^{-1}+q\tilde{D}^{-1}(r_g)\right]^{-1},
\end{align}
where we have introduced~$f(r_g)=1-\int_\mathrm{min}^{r_g}p(l)\dd{l}$ and~$\tilde{D}^{-1}(r_g)=\int_{r_g}^{l_\mathrm{max}}D_1^{-1}(l)\,p(l)\dd{l}$.
We compute the integrals using~$p(l)\sim l^{-\alpha}$ with proper normalization 
as~$f(r_g)=(1-r_g^{1-\alpha}/l_\mathrm{max}^{1-\alpha})$
and~$\tilde{D}^{-1}(r_g)=\frac{1-\alpha}{\alpha}(r_g^{-\alpha}-l_\mathrm{max}^{-\alpha})/l_\mathrm{max}^{1-\alpha}$.
After sorting terms with~$r_g$ and summarizing constants, we arrive at
\begin{equation}
    D^{-1}(r_g)=\frac{1-\alpha}{\alpha \,l_\mathrm{max}}\left(\frac{r_g}{l_\mathrm{max}}\right)^{-\alpha}+\frac{q}{D_0}\left(\frac{r_g}{l_\mathrm{max}}\right)^{1-\alpha}+\mathrm{const},
\end{equation}
which scales as~$D(r_g)\sim r_g^{\alpha}$ for~$r_g\ll l_\mathrm{max}$,
and we recover an energy-dependent diffusion coefficient from energy-independent dynamics.

We stress the strongly simplified nature of our consideration, which merely serves as a proof of concept to show how CR transport could be modified in an inhomogeneous ISM.
While a discussion on the nature of the postulated intermittent patches is beyond the scope of this paper, we note that, in addition to macroscopic structures such as molecular clouds, the effect of dynamo action should be taken into account, which facilitates broad intermittent distributions of~$L_\parallel$,~$\delta B/B_0$ and density \citep[see, e.g.,][]{Gent2024,Beattie2025b}.

\section{Conclusion}\label{sec:conclusion}
We have studied the combined role of magnetic mirroring and resonant curvature scattering in the transport of highly magnetized particles in Alfvénic magnetohydrodynamic turbulence.
We found that these mechanisms facilitate pitch-angle reversals of particles with broad probability distributions of the times between consecutive reversals.
By modeling this process as a Lévy walk, the parallel diffusion coefficient can be derived from these distributions.
While magnetic mirroring plays a prominent role for low-energy particles, the longest involved time scales belong to trajectories with pitch-angle cosines close to one, such that they fall in the loss cones of typical mirroring structures
and eventually scatter off of resonant field line curvature.

Further, we have argued that the scattering mechanisms are intrinsically connected to the magnetic field line structure, where magnetic mirroring occurs in locally ordered field line bundles, while resonant curvature scattering is linked to chaotic regions of the field.
From this point of view, energy-independent parallel transport appears plausible, provided that high-curvature events are clustered, where the sharpest turns are closely surrounded by slightly less intense turns. 
This would lead to highly localized scattering centers that act equally on a wide range of particle energies.
Our results on perpendicular transport also corroborate the role of the field line geometry, where magnetic mirroring diminishes the diffusion coefficients by effectively confining particles to coherent patches, while resonant curvature scattering enhances it by enabling cross-field transport over chaotically separating field lines.

To reconcile our results with energy-dependent observations of galactic cosmic-ray diffusion coefficients, we present a simplified model of an intermittently inhomogeneous interstellar medium, where energy-dependency arises from energy-independent scattering mechanisms via selective interaction of cosmic rays with variously sized turbulent patches.

Although we have focused in this paper on one specific turbulence setup to clarify the roles of the scattering mechanisms at work and their connection to the field line geometry, we note the strong dependency of cosmic-ray transport coefficients on turbulence parameters (such as Alfvén and sonic Mach numbers) and driving mechanisms, which warrants more detailed studies in the future.

\begin{acknowledgments}
J.L. gratefully acknowledges helpful discussions with
Sophie Aerdker,
James R.~Beattie,
Robert J.~Ewart,
Martin Lemoine,
Patrick Reichherzer
and Siyao Xu.
This work was supported by
the Deutsche Forschungsgemeinschaft (DFG, German Research Foundation) through the Collaborative Research Center SFB1491 ``Cosmic Interacting Matters - From Source to Signal'' (grant no.~445052434);
and the International Space Science Institute (ISSI) in Bern through ISSI International Team project \#24-608 ``Energetic Particle Transport in Space Plasma Turbulence''.
The authors gratefully acknowledge the Gauss Centre for Supercomputing e.V. (\url{www.gauss-centre.eu}) for funding this project by providing computing time on the SuperMUC-NG at Leibniz Supercomputing Centre (\url{www.lrz.de}) and through the John von Neumann Institute for Computing (NIC) on the GCS Supercomputers JUWELS at Jülich Supercomputing Centre (JSC).
\end{acknowledgments}

\appendix
\section{MHD Simulation}\label{app:mhd}
We simulate incompressible visco-resistive MHD turbulence with our pseudo-spectral code \textit{SpecDyn} \citep{Wilbert2022,WilbertPhd2023} on an elongated box, measuring~$L_x,L_y=2\pi$ and~$L_z=\sqrt{2}\times2\pi$, with periodic boundary conditions, resolved on a grid with~$1024^3$ points.
The governing equations for the flow~$\vb{u}$ and magnetic field~$\vb{B}$ are given by
\begin{subequations}
    \begin{gather}
        \frac{\partial\vb{u}}{\partial t} +\vb{u} \bcdot \bnabla \vb{u} = \vb{B} \bcdot \bnabla \vb{B}
        -\bnabla p +\nu\Delta \vb{u} +\vb{f}, \quad \bnabla\bcdot\vb{u} = 0, \\
        \frac{\partial\vb{B}}{\partial t}+\vb{u}\bcdot\bnabla\vb{B} -\vb{B}\bcdot\bnabla\vb{u} =
        \eta\Delta \vb{B}, \quad \bnabla\bcdot\vb{B} = 0.
    \end{gather}
\end{subequations}

The magnetic field~$\vb{B}=\vb{B}_0+\delta\vb{B}$ consists of a constant uniform background component~$\vb{B}_0=B_0\hat{\vb{z}}$ and a fluctuating component~$\delta\vb{B}$.
The fields~$\vb{u}$ and~$\delta\vb{B}$ are initialized with small-amplitude random fields with zero total cross- and magnetic helicity.
The flow~$\vb{u}$ is driven by the force density~$\vb{f}$ on the wave number shell~$1\leq k\leq 2$ with random amplitudes, zero cross-helicity injection and~$\delta$-correlation in time to ensure constant power injection.
The values for viscosity and resistivity are chosen equal as~$\nu=\eta=1.87\times10^{-3}$, resulting in Taylor-scale Reynolds numbers~$Re_{u,T}=261.744$ and~$Re_{B,T}=178.698$.
We run the simulation until energy and dissipation rate are statistically saturated and then extract 8 snapshots separated in time by~$T_\mathrm{eddy}=L_x/u_\mathrm{rms}$.

\bibliographystyle{aasjournalv7}
\bibliography{main}

\begin{thebibliography}{}
\expandafter\ifx\csname natexlab\endcsname\relax\def\natexlab#1{#1}\fi
\providecommand{\url}[1]{\href{#1}{#1}}
\providecommand{\dodoi}[1]{doi:~\href{http://doi.org/#1}{\nolinkurl{#1}}}
\providecommand{\doeprint}[1]{\href{http://ascl.net/#1}{\nolinkurl{http://ascl.net/#1}}}
\providecommand{\doarXiv}[1]{\href{https://arxiv.org/abs/#1}{\nolinkurl{https://arxiv.org/abs/#1}}}

\bibitem[{S. {Aerdker} {et~al.}(2025){Aerdker}, {Merten}, {Effenberger}, {Fichtner}, \& {Becker Tjus}}]{Aerdker2025}
{Aerdker}, S., {Merten}, L., {Effenberger}, F., {Fichtner}, H., \& {Becker Tjus}, J. 2025, \bibinfo{title}{{Superdiffusion of energetic particles at shocks: A L{\'e}vy flight model for acceleration},} \aap, 693, A15, \dodoi{10.1051/0004-6361/202451765}

\bibitem[{M. {Aguilar} {et~al.}(2016){Aguilar}, {Ali Cavasonza}, {Ambrosi}, {Arruda}, {Attig}, {Aupetit}, {Azzarello}, {Bachlechner}, {Barao}, {Barrau}, {Barrin}, {Bartoloni}, {Basara}, {Ba{\c{s}}e{\v{g}}mez-du Pree}, {Battarbee}, {Battiston}, {Becker}, {Behlmann}, {Beischer}, {Berdugo}, {Bertucci}, {Bindel}, {Bindi}, {Boella}, {de Boer}, {Bollweg}, {Bonnivard}, {Borgia}, {Boschini}, {Bourquin}, {Bueno}, {Burger}, {Cadoux}, {Cai}, {Capell}, {Caroff}, {Casaus}, {Castellini}, {Cervelli}, {Chae}, {Chang}, {Chen}, {Chen}, {Chen}, {Cheng}, {Chou}, {Choumilov}, {Choutko}, {Chung}, {Clark}, {Clavero}, {Coignet}, {Consolandi}, {Contin}, {Corti}, {Creus}, {Crispoltoni}, {Cui}, {Dai}, {Delgado}, {Della Torre}, {Demakov}, {Demirk{\"o}z}, {Derome}, {Di Falco}, {Dimiccoli}, {D{\'\i}az}, {von Doetinchem}, {Dong}, {Donnini}, {Duranti}, {D'Urso}, {Egorov}, {Eline}, {Eronen}, {Feng}, {Fiandrini}, {Finch}, {Fisher}, {Formato}, {Galaktionov}, {Gallucci}, {Garc{\'\i}a}, {Garc{\'\i}a-L{\'o}pez}, {Gargiulo}, {Gast}, {Gebauer},
  {Gervasi}, {Ghelfi}, {Giovacchini}, {Goglov}, {G{\'o}mez-Coral}, {Gong}, {Goy}, {Grabski}, {Grandi}, {Graziani}, {Guo}, {Haino}, {Han}, {He}, {Heil}, {Hoffman}, {Hsieh}, {Huang}, {Huang}, {Huh}, {Incagli}, {Ionica}, {Jang}, {Jinchi}, {Kang}, {Kanishev}, {Kim}, {Kim}, {Kirn}, {Konak}, {Kounina}, {Kounine}, {Koutsenko}, {Krafczyk}, {La Vacca}, {Laudi}, {Laurenti}, {Lazzizzera}, {Lebedev}, {Lee}, {Lee}, {Leluc}, {Li}, {Li}, {Li}, {Li}, {Li}, {Li}, {Li}, {Li}, {Li}, {Lim}, {Lin}, {Lipari}, {Lippert}, {Liu}, {Liu}, {Lordello}, {Lu}, {Lu}, {Luebelsmeyer}, {Luo}, {Luo}, {Lv}, {Machate}, {Majka}, {Ma{\~n}{\'a}}, {Mar{\'\i}n}, {Martin}, {Mart{\'\i}nez}, {Masi}, {Maurin}, {Menchaca-Rocha}, {Meng}, {Mikuni}, {Mo}, {Morescalchi}, {Mott}, {Nelson}, {Ni}, {Nikonov}, {Nozzoli}, {Oliva}, {Orcinha}, {Palmonari}, {Palomares}, {Paniccia}, {Pauluzzi}, {Pensotti}, {Pereira}, {Picot-Clemente}, {Pilo}, {Pizzolotto}, {Plyaskin}, {Pohl}, {Poireau}, {Putze}, {Quadrani}, {Qi}, {Qin}, {Qu}, {R{\"a}ih{\"a}}, {Rancoita}, {Rapin},
  {Ricol}, {Rosier-Lees}, {Rozhkov}, {Rozza}, {Sagdeev}, {Sandweiss}, {Saouter}, {Schael}, \& {Schmidt}}]{Aguilar2016}
{Aguilar}, M., {Ali Cavasonza}, L., {Ambrosi}, G., {et~al.} 2016, \bibinfo{title}{{Precision Measurement of the Boron to Carbon Flux Ratio in Cosmic Rays from 1.9 GV to 2.6 TV with the Alpha Magnetic Spectrometer on the International Space Station},} \prl, 117, 231102, \dodoi{10.1103/PhysRevLett.117.231102}

\bibitem[{B.~J. {Albright} {et~al.}(2001){Albright}, {Chandran}, {Cowley}, \& {Loh}}]{Albright2001}
{Albright}, B.~J., {Chandran}, B.~D.~G., {Cowley}, S.~C., \& {Loh}, M. 2001, \bibinfo{title}{{Parallel heat diffusion and subdiffusion in random magnetic fields},} Physics of Plasmas, 8, 777, \dodoi{10.1063/1.1344920}

\bibitem[{E. {Amato} \& P. {Blasi}(2018){Amato} \& {Blasi}}]{Amato2018}
{Amato}, E., \& {Blasi}, P. 2018, \bibinfo{title}{{Cosmic ray transport in the Galaxy: A review},} Advances in Space Research, 62, 2731, \dodoi{10.1016/j.asr.2017.04.019}

\bibitem[{R. {Bandyopadhyay} {et~al.}(2020){Bandyopadhyay}, {Yang}, {Matthaeus}, {Chasapis}, {Parashar}, {Russell}, {Strangeway}, {Torbert}, {Giles}, {Gershman}, {Pollock}, {Moore}, \& {Burch}}]{Bandyopadhyay2020}
{Bandyopadhyay}, R., {Yang}, Y., {Matthaeus}, W.~H., {et~al.} 2020, \bibinfo{title}{{In Situ Measurement of Curvature of Magnetic Field in Turbulent Space Plasmas: A Statistical Study},} \apjl, 893, L25, \dodoi{10.3847/2041-8213/ab846e}

\bibitem[{L. {Barreto-Mota} {et~al.}(2025){Barreto-Mota}, {de Gouveia Dal Pino}, {Xu}, \& {Lazarian}}]{Barreto-Mota2025}
{Barreto-Mota}, L., {de Gouveia Dal Pino}, E.~M., {Xu}, S., \& {Lazarian}, A. 2025, \bibinfo{title}{{Cosmic-Ray Diffusion in the Turbulent Interstellar Medium: Effects of Mirror Diffusion and Pitch-angle Scattering},} \apj, 988, 269, \dodoi{10.3847/1538-4357/ade4c8}

\bibitem[{J.~R. {Beattie} {et~al.}(2024){Beattie}, {Federrath}, {Klessen}, {Cielo}, \& {Bhattacharjee}}]{Beattie2024}
{Beattie}, J.~R., {Federrath}, C., {Klessen}, R.~S., {Cielo}, S., \& {Bhattacharjee}, A. 2024, \bibinfo{title}{{Magnetized compressible turbulence with a fluctuation dynamo and Reynolds numbers over a million},} arXiv e-prints, arXiv:2405.16626, \dodoi{10.48550/arXiv.2405.16626}

\bibitem[{J.~R. {Beattie} {et~al.}(2025){Beattie}, {Noer Kolborg}, {Ramirez-Ruiz}, \& {Federrath}}]{Beattie2025b}
{Beattie}, J.~R., {Noer Kolborg}, A., {Ramirez-Ruiz}, E., \& {Federrath}, C. 2025, \bibinfo{title}{{So long Kolmogorov: the forward and backward turbulence cascades in a supernovae-driven, multiphase interstellar medium},} arXiv e-prints, arXiv:2501.09855, \dodoi{10.48550/arXiv.2501.09855}

\bibitem[{A.~R. {Bell}(2013){Bell}}]{Bell2013}
{Bell}, A.~R. 2013, \bibinfo{title}{{Cosmic ray acceleration},} Astroparticle Physics, 43, 56, \dodoi{10.1016/j.astropartphys.2012.05.022}

\bibitem[{A.~R. {Bell} {et~al.}(2025){Bell}, {Matthews}, {Taylor}, \& {Giacinti}}]{Bell2025}
{Bell}, A.~R., {Matthews}, J.~H., {Taylor}, A.~M., \& {Giacinti}, G. 2025, \bibinfo{title}{{Cosmic ray transport and acceleration with magnetic mirroring},} \mnras, 539, 1236, \dodoi{10.1093/mnras/staf562}

\bibitem[{A. {Beresnyak} {et~al.}(2011){Beresnyak}, {Yan}, \& {Lazarian}}]{Beresnyak2011}
{Beresnyak}, A., {Yan}, H., \& {Lazarian}, A. 2011, \bibinfo{title}{{Numerical Study of Cosmic Ray Diffusion in Magnetohydrodynamic Turbulence},} \apj, 728, 60, \dodoi{10.1088/0004-637X/728/1/60}

\bibitem[{J.~P. Boris \& R.~A. Shanny(1971)Boris \& Shanny}]{boris:1971}
Boris, J.~P., \& Shanny, R.~A. 1971, Proceedings, Fourth Conference on Numerical Simulation of Plasmas (Naval Research Laboratory)

\bibitem[{I.~S. {Butsky} {et~al.}(2024){Butsky}, {Hopkins}, {Kempski}, {Ponnada}, {Quataert}, \& {Squire}}]{Butsky2024}
{Butsky}, I.~S., {Hopkins}, P.~F., {Kempski}, P., {et~al.} 2024, \bibinfo{title}{{Galactic cosmic-ray scattering due to intermittent structures},} \mnras, 528, 4245, \dodoi{10.1093/mnras/stae276}

\bibitem[{B.~D.~G. {Chandran}(2000{\natexlab{a}}){Chandran}}]{Chandran2000a}
{Chandran}, B. D.~G. 2000{\natexlab{a}}, \bibinfo{title}{{Scattering of Energetic Particles by Anisotropic Magnetohydrodynamic Turbulence with a Goldreich-Sridhar Power Spectrum},} \prl, 85, 4656, \dodoi{10.1103/PhysRevLett.85.4656}

\bibitem[{B.~D.~G. {Chandran}(2000{\natexlab{b}}){Chandran}}]{Chandran2000b}
{Chandran}, B. D.~G. 2000{\natexlab{b}}, \bibinfo{title}{{Confinement and Isotropization of Galactic Cosmic Rays by Molecular-Cloud Magnetic Mirrors When Turbulent Scattering Is Weak},} \apj, 529, 513, \dodoi{10.1086/308232}

\bibitem[{F.~F. {Chen}(2016){Chen}}]{Chen2016_with_publisher}
{Chen}, F.~F. 2016, {Introduction to Plasma Physics and Controlled Fusion} (Springer International Publishing), \dodoi{10.1007/978-3-319-22309-4}

\bibitem[{R.~M. {Crutcher}(1999){Crutcher}}]{Crutcher1999}
{Crutcher}, R.~M. 1999, \bibinfo{title}{{Magnetic Fields in Molecular Clouds: Observations Confront Theory},} \apj, 520, 706, \dodoi{10.1086/307483}

\bibitem[{P. {Desiati} \& E.~G. {Zweibel}(2014){Desiati} \& {Zweibel}}]{Desiati2014}
{Desiati}, P., \& {Zweibel}, E.~G. 2014, \bibinfo{title}{{The Transport of Cosmic Rays Across Magnetic Fieldlines},} \apj, 791, 51, \dodoi{10.1088/0004-637X/791/1/51}

\bibitem[{C. {Dong} {et~al.}(2022){Dong}, {Wang}, {Huang}, {Comisso}, {Sandstrom}, \& {Bhattacharjee}}]{Dong2022}
{Dong}, C., {Wang}, L., {Huang}, Y.-M., {et~al.} 2022, \bibinfo{title}{{Reconnection-driven energy cascade in magnetohydrodynamic turbulence},} Science Advances, 8, eabn7627, \dodoi{10.1126/sciadv.abn7627}

\bibitem[{J. {D{\"o}rner} {et~al.}(2024){D{\"o}rner}, {Becker Tjus}, {Blomenkamp}, {Fichtner}, {Franckowiak}, \& {Zaninger}}]{Dorner2024}
{D{\"o}rner}, J., {Becker Tjus}, J., {Blomenkamp}, P.~S., {et~al.} 2024, \bibinfo{title}{{Impact of Anisotropic Cosmic-Ray Transport on the Gamma-Ray Signatures in the Galactic Center},} \apj, 965, 180, \dodoi{10.3847/1538-4357/ad2ea1}

\bibitem[{F. {Effenberger} {et~al.}(2012){Effenberger}, {Fichtner}, {Scherer}, \& {B{\"u}sching}}]{Effenberger2012a}
{Effenberger}, F., {Fichtner}, H., {Scherer}, K., \& {B{\"u}sching}, I. 2012, \bibinfo{title}{{Anisotropic diffusion of Galactic cosmic ray protons and their steady-state azimuthal distribution},} \aap, 547, A120, \dodoi{10.1051/0004-6361/201220203}

\bibitem[{N.~E. {Engelbrecht} {et~al.}(2022){Engelbrecht}, {Effenberger}, {Florinski}, {Potgieter}, {Ruffolo}, {Chhiber}, {Usmanov}, {Rankin}, \& {Els}}]{Engelbrecht2022}
{Engelbrecht}, N.~E., {Effenberger}, F., {Florinski}, V., {et~al.} 2022, \bibinfo{title}{{Theory of Cosmic Ray Transport in the Heliosphere},} \ssr, 218, 33, \dodoi{10.1007/s11214-022-00896-1}

\bibitem[{R.~J. {Ewart} {et~al.}(2025){Ewart}, {Reichherzer}, {Ren}, {Majeski}, {Mori}, {Nastac}, {Bott}, {Kunz}, \& {Schekochihin}}]{Ewart2025}
{Ewart}, R.~J., {Reichherzer}, P., {Ren}, S., {et~al.} 2025, \bibinfo{title}{{Cosmic-ray transport in inhomogeneous media},} arXiv e-prints, arXiv:2507.19044, \dodoi{10.48550/arXiv.2507.19044}

\bibitem[{O. {Fornieri} {et~al.}(2021){Fornieri}, {Gaggero}, {Cerri}, {De La Torre Luque}, \& {Gabici}}]{Fornieri2021}
{Fornieri}, O., {Gaggero}, D., {Cerri}, S.~S., {De La Torre Luque}, P., \& {Gabici}, S. 2021, \bibinfo{title}{{The theory of cosmic ray scattering on pre-existing MHD modes meets data},} \mnras, 502, 5821, \dodoi{10.1093/mnras/stab355}

\bibitem[{F.~A. {Gent} {et~al.}(2024){Gent}, {Mac Low}, \& {Korpi-Lagg}}]{Gent2024}
{Gent}, F.~A., {Mac Low}, M.-M., \& {Korpi-Lagg}, M.~J. 2024, \bibinfo{title}{{Transition from Small-scale to Large-scale Dynamo in a Supernova-driven, Multiphase Medium},} \apj, 961, 7, \dodoi{10.3847/1538-4357/ad0da0}

\bibitem[{P. {Goldreich} \& S. {Sridhar}(1995){Goldreich} \& {Sridhar}}]{Goldreich1995}
{Goldreich}, P., \& {Sridhar}, S. 1995, \bibinfo{title}{{Toward a Theory of Interstellar Turbulence. II. Strong Alfvenic Turbulence},} \apj, 438, 763, \dodoi{10.1086/175121}

\bibitem[{R. {Grauer} {et~al.}(1994){Grauer}, {Krug}, \& {Marliani}}]{Grauer1994}
{Grauer}, R., {Krug}, J., \& {Marliani}, C. 1994, \bibinfo{title}{{Scaling of high-order structure functions in magnetohydrodynamic turbulence},} Physics Letters A, 195, 335, \dodoi{10.1016/0375-9601(94)90038-8}

\bibitem[{P.~C. {Gray} \& L.~C. {Lee}(1982){Gray} \& {Lee}}]{Gray1982}
{Gray}, P.~C., \& {Lee}, L.~C. 1982, \bibinfo{title}{{Particle pitch angle diffusion due to nonadiabatic effects in the plasma sheet},} \jgr, 87, 7445, \dodoi{10.1029/JA087iA09p07445}

\bibitem[{A. {Greco} {et~al.}(2008){Greco}, {Chuychai}, {Matthaeus}, {Servidio}, \& {Dmitruk}}]{Greco2008}
{Greco}, A., {Chuychai}, P., {Matthaeus}, W.~H., {Servidio}, S., \& {Dmitruk}, P. 2008, \bibinfo{title}{{Intermittent MHD structures and classical discontinuities},} \grl, 35, L19111, \dodoi{10.1029/2008GL035454}

\bibitem[{V. {Heesen} {et~al.}(2023){Heesen}, {de Gasperin}, {Schulz}, {Basu}, {Beck}, {Br{\"u}ggen}, {Dettmar}, {Stein}, {Gajovi{\'c}}, {Tabatabaei}, \& {Reichherzer}}]{Heesen2023}
{Heesen}, V., {de Gasperin}, F., {Schulz}, S., {et~al.} 2023, \bibinfo{title}{{Diffusion of cosmic-ray electrons in M 51 observed with LOFAR at 54 MHz},} \aap, 672, A21, \dodoi{10.1051/0004-6361/202245223}

\bibitem[{P.~F. {Hopkins} {et~al.}(2022){Hopkins}, {Squire}, {Butsky}, \& {Ji}}]{Hopkins2022a}
{Hopkins}, P.~F., {Squire}, J., {Butsky}, I.~S., \& {Ji}, S. 2022, \bibinfo{title}{{Standard self-confinement and extrinsic turbulence models for cosmic ray transport are fundamentally incompatible with observations},} \mnras, 517, 5413, \dodoi{10.1093/mnras/stac2909}

\bibitem[{Y. {Hu} {et~al.}(2022){Hu}, {Lazarian}, \& {Xu}}]{Hu2022}
{Hu}, Y., {Lazarian}, A., \& {Xu}, S. 2022, \bibinfo{title}{{Superdiffusion of cosmic rays in compressible magnetized turbulence},} \mnras, 512, 2111, \dodoi{10.1093/mnras/stac319}

\bibitem[{Y.-M. {Huang} {et~al.}(2014){Huang}, {Bhattacharjee}, \& {Boozer}}]{Huang2014}
{Huang}, Y.-M., {Bhattacharjee}, A., \& {Boozer}, A.~H. 2014, \bibinfo{title}{{Rapid Change of Field Line Connectivity and Reconnection in Stochastic Magnetic Fields},} \apj, 793, 106, \dodoi{10.1088/0004-637X/793/2/106}

\bibitem[{J.~R. {Jokipii}(1966){Jokipii}}]{Jokipii1966}
{Jokipii}, J.~R. 1966, \bibinfo{title}{{Cosmic-Ray Propagation. I. Charged Particles in a Random Magnetic Field},} \apj, 146, 480, \dodoi{10.1086/148912}

\bibitem[{P. {Kempski} {et~al.}(2025){Kempski}, {Fielding}, {Quataert}, {Ewart}, {Grete}, {Kunz}, {Philippov}, \& {Stone}}]{Kempski2025b}
{Kempski}, P., {Fielding}, D.~B., {Quataert}, E., {et~al.} 2025, \bibinfo{title}{{Self-Similar Cosmic-Ray Transport in High-Resolution Magnetohydrodynamic Turbulence},} arXiv e-prints, arXiv:2507.10651, \dodoi{10.48550/arXiv.2507.10651}

\bibitem[{P. {Kempski} {et~al.}(2023){Kempski}, {Fielding}, {Quataert}, {Galishnikova}, {Kunz}, {Philippov}, \& {Ripperda}}]{Kempski2023}
{Kempski}, P., {Fielding}, D.~B., {Quataert}, E., {et~al.} 2023, \bibinfo{title}{{Cosmic ray transport in large-amplitude turbulence with small-scale field reversals},} \mnras, 525, 4985, \dodoi{10.1093/mnras/stad2609}

\bibitem[{P. {Kempski} \& E. {Quataert}(2022){Kempski} \& {Quataert}}]{Kempski2022}
{Kempski}, P., \& {Quataert}, E. 2022, \bibinfo{title}{{Reconciling cosmic ray transport theory with phenomenological models motivated by Milky-Way data},} \mnras, 514, 657, \dodoi{10.1093/mnras/stac1240}

\bibitem[{J. {Kleimann} {et~al.}(2025){Kleimann}, {Fichtner}, {Stein}, {Dettmar}, {Bomans}, \& {Oughton}}]{Kleimann2025}
{Kleimann}, J., {Fichtner}, H., {Stein}, M., {et~al.} 2025, \bibinfo{title}{{Magnetohydrodynamic turbulence and the associated spatial diffusion tensor of cosmic rays in dynamical galactic halos},} \aap, 699, A92, \dodoi{10.1051/0004-6361/202553873}

\bibitem[{N. {Kriel} {et~al.}(2025){Kriel}, {Beattie}, {Federrath}, {Krumholz}, \& {Hew}}]{Kriel2025}
{Kriel}, N., {Beattie}, J.~R., {Federrath}, C., {Krumholz}, M.~R., \& {Hew}, J. K.~J. 2025, \bibinfo{title}{{Fundamental MHD scales - II. The kinematic phase of the supersonic small-scale dynamo},} \mnras, 537, 2602, \dodoi{10.1093/mnras/staf188}

\bibitem[{R. {Kulsrud} \& W.~P. {Pearce}(1969){Kulsrud} \& {Pearce}}]{Kulsrud1969}
{Kulsrud}, R., \& {Pearce}, W.~P. 1969, \bibinfo{title}{{The Effect of Wave-Particle Interactions on the Propagation of Cosmic Rays},} \apj, 156, 445, \dodoi{10.1086/149981}

\bibitem[{T. {Laitinen} \& S. {Dalla}(2017){Laitinen} \& {Dalla}}]{Laitinen2017b}
{Laitinen}, T., \& {Dalla}, S. 2017, \bibinfo{title}{{Energetic Particle Transport across the Mean Magnetic Field: Before Diffusion},} \apj, 834, 127, \dodoi{10.3847/1538-4357/834/2/127}

\bibitem[{A. {Lazarian} \& S. {Xu}(2021){Lazarian} \& {Xu}}]{Lazarian2021}
{Lazarian}, A., \& {Xu}, S. 2021, \bibinfo{title}{{Diffusion of Cosmic Rays in MHD Turbulence with Magnetic Mirrors},} \apj, 923, 53, \dodoi{10.3847/1538-4357/ac2de9}

\bibitem[{M. {Lemoine}(2023){Lemoine}}]{Lemoine2023}
{Lemoine}, M. 2023, \bibinfo{title}{{Particle transport through localized interactions with sharp magnetic field bends in MHD turbulence},} Journal of Plasma Physics, 89, 175890501, \dodoi{10.1017/S0022377823000946}

\bibitem[{N. {Liang} \& S.~P. {Oh}(2025){Liang} \& {Oh}}]{Liang2025}
{Liang}, N., \& {Oh}, S.~P. 2025, \bibinfo{title}{{L{\'e}vy flights and leaky boxes: anomalous diffusion of cosmic rays},} \mnras, 543, 1911, \dodoi{10.1093/mnras/staf1474}

\bibitem[{V. {L{\'o}pez-Barquero} \& P. {Desiati}(2025){L{\'o}pez-Barquero} \& {Desiati}}]{Lopez-Barquero2025}
{L{\'o}pez-Barquero}, V., \& {Desiati}, P. 2025, \bibinfo{title}{{Chaotic Behavior of Trapped Cosmic Rays},} \apj, 983, 106, \dodoi{10.3847/1538-4357/adbca7}

\bibitem[{J. {L{\"u}bke} {et~al.}(2024){L{\"u}bke}, {Effenberger}, {Wilbert}, {Fichtner}, \& {Grauer}}]{Lubke2024}
{L{\"u}bke}, J., {Effenberger}, F., {Wilbert}, M., {Fichtner}, H., \& {Grauer}, R. 2024, \bibinfo{title}{{Towards synthetic magnetic turbulence with coherent structures},} EPL (Europhysics Letters), 146, 43001, \dodoi{10.1209/0295-5075/ad438f}

\bibitem[{J. {L{\"u}bke} {et~al.}(2025){L{\"u}bke}, {Reichherzer}, {Aerdker}, {Effenberger}, {Wilbert}, {Fichtner}, \& {Grauer}}]{Lubke2025b}
{L{\"u}bke}, J., {Reichherzer}, P., {Aerdker}, S., {et~al.} 2025, \bibinfo{title}{{Modelling Cosmic-Ray Transport: Magnetized versus Unmagnetized Motion in Astrophysical Magnetic Turbulence},} arXiv e-prints, arXiv:2505.18155, \dodoi{10.48550/arXiv.2505.18155}

\bibitem[{M. {Magdziarz} \& A. {Weron}(2007){Magdziarz} \& {Weron}}]{Magdziarz2007}
{Magdziarz}, M., \& {Weron}, A. 2007, \bibinfo{title}{{Competition between subdiffusion and L{\'e}vy flights: A Monte Carlo approach},} \pre, 75, 056702, \dodoi{10.1103/PhysRevE.75.056702}

\bibitem[{F. {Malara} {et~al.}(2023){Malara}, {Perri}, {Giacalone}, \& {Zimbardo}}]{Malara2023}
{Malara}, F., {Perri}, S., {Giacalone}, J., \& {Zimbardo}, G. 2023, \bibinfo{title}{{Energetic particle dynamics in a simplified model of a solar wind magnetic switchback},} \aap, 677, A69, \dodoi{10.1051/0004-6361/202346990}

\bibitem[{A. {Mallet} {et~al.}(2015){Mallet}, {Schekochihin}, \& {Chandran}}]{Mallet2015}
{Mallet}, A., {Schekochihin}, A.~A., \& {Chandran}, B.~D.~G. 2015, \bibinfo{title}{{Refined critical balance in strong Alfvenic turbulence.},} \mnras, 449, L77, \dodoi{10.1093/mnrasl/slv021}

\bibitem[{F. {Manke} {et~al.}(2019){Manke}, {Baquero-Ruiz}, {Furno}, {Chella{\"\i}}, {Fasoli}, \& {Ricci}}]{Manke2019}
{Manke}, F., {Baquero-Ruiz}, M., {Furno}, I., {et~al.} 2019, \bibinfo{title}{{Truncated L{\'e}vy motion through path integrals and applications to nondiffusive suprathermal ion transport},} \pre, 100, 052122, \dodoi{10.1103/PhysRevE.100.052122}

\bibitem[{J. {Martin} {et~al.}(2025){Martin}, {L{\"u}bke}, {Li}, {Buzzicotti}, {Grauer}, \& {Biferale}}]{Martin2025}
{Martin}, J., {L{\"u}bke}, J., {Li}, T., {et~al.} 2025, \bibinfo{title}{{Generation of Cosmic-Ray Trajectories by a Diffusion Model Trained on Test Particles in 3D Magnetohydrodynamic Turbulence},} \apjs, 277, 48, \dodoi{10.3847/1538-4365/adb432}

\bibitem[{P. {Mertsch}(2020){Mertsch}}]{Mertsch2020}
{Mertsch}, P. 2020, \bibinfo{title}{{Test particle simulations of cosmic rays},} \apss, 365, 135, \dodoi{10.1007/s10509-020-03832-3}

\bibitem[{H. Mori(1965)Mori}]{mori:1965}
Mori, H. 1965, \bibinfo{title}{Transport, Collective Motion, and Brownian Motion*),} Progress of Theoretical Physics, 33, 423, \dodoi{10.1143/PTP.33.423}

\bibitem[{P.~D. {Noerdlinger}(1968){Noerdlinger}}]{Noerdlinger1968}
{Noerdlinger}, P.~D. 1968, \bibinfo{title}{{An Improved Model for Cosmic-Ray Propagation},} \prl, 20, 1513, \dodoi{10.1103/PhysRevLett.20.1513}

\bibitem[{E. {Ntormousi} {et~al.}(2024){Ntormousi}, {Vlahos}, {Konstantinou}, \& {Isliker}}]{Ntormousi2024}
{Ntormousi}, E., {Vlahos}, L., {Konstantinou}, A., \& {Isliker}, H. 2024, \bibinfo{title}{{Strong turbulence and magnetic coherent structures in the interstellar medium},} \aap, 691, A149, \dodoi{10.1051/0004-6361/202450710}

\bibitem[{O. {Pezzi} \& P. {Blasi}(2024){Pezzi} \& {Blasi}}]{Pezzi2024}
{Pezzi}, O., \& {Blasi}, P. 2024, \bibinfo{title}{{Galactic cosmic ray transport in the absence of resonant scattering},} \mnras, 529, L13, \dodoi{10.1093/mnrasl/slad192}

\bibitem[{O. {Pezzi} {et~al.}(2022){Pezzi}, {Blasi}, \& {Matthaeus}}]{Pezzi2022}
{Pezzi}, O., {Blasi}, P., \& {Matthaeus}, W.~H. 2022, \bibinfo{title}{{Relativistic Particle Transport and Acceleration in Structured Plasma Turbulence},} \apj, 928, 25, \dodoi{10.3847/1538-4357/ac5332}

\bibitem[{ {Planck Collaboration} {et~al.}(2016){Planck Collaboration}, {Ade}, {Aghanim}, {Alves}, {Arnaud}, {Arzoumanian}, {Ashdown}, {Aumont}, {Baccigalupi}, {Banday}, {Barreiro}, {Bartolo}, {Battaner}, {Benabed}, {Beno{\^\i}t}, {Benoit-L{\'e}vy}, {Bernard}, {Bersanelli}, {Bielewicz}, {Bock}, {Bonavera}, {Bond}, {Borrill}, {Bouchet}, {Boulanger}, {Bracco}, {Burigana}, {Calabrese}, {Cardoso}, {Catalano}, {Chiang}, {Christensen}, {Colombo}, {Combet}, {Couchot}, {Crill}, {Curto}, {Cuttaia}, {Danese}, {Davies}, {Davis}, {de Bernardis}, {de Rosa}, {de Zotti}, {Delabrouille}, {Dickinson}, {Diego}, {Dole}, {Donzelli}, {Dor{\'e}}, {Douspis}, {Ducout}, {Dupac}, {Efstathiou}, {Elsner}, {En{\ss}lin}, {Eriksen}, {Falceta-Gon{\c{c}}alves}, {Falgarone}, {Ferri{\`e}re}, {Finelli}, {Forni}, {Frailis}, {Fraisse}, {Franceschi}, {Frejsel}, {Galeotta}, {Galli}, {Ganga}, {Ghosh}, {Giard}, {Gjerl{\o}w}, {Gonz{\'a}lez-Nuevo}, {G{\'o}rski}, {Gregorio}, {Gruppuso}, {Gudmundsson}, {Guillet}, {Harrison}, {Helou}, {Hennebelle},
  {Henrot-Versill{\'e}}, {Hern{\'a}ndez-Monteagudo}, {Herranz}, {Hildebrandt}, {Hivon}, {Holmes}, {Hornstrup}, {Huffenberger}, {Hurier}, {Jaffe}, {Jaffe}, {Jones}, {Juvela}, {Keih{\"a}nen}, {Keskitalo}, {Kisner}, {Knoche}, {Kunz}, {Kurki-Suonio}, {Lagache}, {Lamarre}, {Lasenby}, {Lattanzi}, {Lawrence}, {Leonardi}, {Levrier}, {Liguori}, {Lilje}, {Linden-V{\o}rnle}, {L{\'o}pez-Caniego}, {Lubin}, {Mac{\'\i}as-P{\'e}rez}, {Maino}, {Mandolesi}, {Mangilli}, {Maris}, {Martin}, {Mart{\'\i}nez-Gonz{\'a}lez}, {Masi}, {Matarrese}, {Melchiorri}, {Mendes}, {Mennella}, {Migliaccio}, {Miville-Desch{\^e}nes}, {Moneti}, {Montier}, {Morgante}, {Mortlock}, {Munshi}, {Murphy}, {Naselsky}, {Nati}, {Netterfield}, {Noviello}, {Novikov}, {Novikov}, {Oppermann}, {Oxborrow}, {Pagano}, {Pajot}, {Paladini}, {Paoletti}, {Pasian}, {Perotto}, {Pettorino}, {Piacentini}, {Piat}, {Pierpaoli}, {Pietrobon}, {Plaszczynski}, {Pointecouteau}, {Polenta}, {Ponthieu}, {Pratt}, {Prunet}, {Puget}, {Rachen}, {Reinecke}, {Remazeilles}, {Renault},
  {Renzi}, {Ristorcelli}, {Rocha}, {Rossetti}, {Roudier}, {Rubi{\~n}o-Mart{\'\i}n}, {Rusholme}, {Sandri}, {Santos}, {Savelainen}, {Savini}, {Scott}, {Soler}, {Stolyarov}, {Sudiwala}, {Sutton}, {Suur-Uski}, {Sygnet}, {Tauber}, {Terenzi}, {Toffolatti}, {Tomasi}, {Tristram}, {Tucci}, {Umana}, {Valenziano}, {Valiviita}, {Van Tent}, {Vielva}, {Villa}, {Wade}, {Wandelt}, {Wehus}, {Ysard}, {Yvon}, \& {Zonca}}]{PlanckCollaboration2016}
{Planck Collaboration}, {Ade}, P.~A.~R., {Aghanim}, N., {et~al.} 2016, \bibinfo{title}{{Planck intermediate results. XXXV. Probing the role of the magnetic field in the formation of structure in molecular clouds},} \aap, 586, A138, \dodoi{10.1051/0004-6361/201525896}

\bibitem[{F. {Pucci} {et~al.}(2016){Pucci}, {Malara}, {Perri}, {Zimbardo}, {Sorriso-Valvo}, \& {Valentini}}]{Pucci2016}
{Pucci}, F., {Malara}, F., {Perri}, S., {et~al.} 2016, \bibinfo{title}{{Energetic particle transport in the presence of magnetic turbulence: influence of spectral extension and intermittency},} \mnras, 459, 3395, \dodoi{10.1093/mnras/stw877}

\bibitem[{P. {Reichherzer} {et~al.}(2025){Reichherzer}, {Bott}, {Ewart}, {Gregori}, {Kempski}, {Kunz}, \& {Schekochihin}}]{Reichherzer2025}
{Reichherzer}, P., {Bott}, A. F.~A., {Ewart}, R.~J., {et~al.} 2025, \bibinfo{title}{{Efficient micromirror confinement of sub-teraelectronvolt cosmic rays in galaxy clusters},} Nature Astronomy, 9, 438, \dodoi{10.1038/s41550-024-02442-1}

\bibitem[{P. {Reichherzer} {et~al.}(2022){Reichherzer}, {Merten}, {D{\"o}rner}, {Becker Tjus}, {Pueschel}, \& {Zweibel}}]{Reichherzer2022b}
{Reichherzer}, P., {Merten}, L., {D{\"o}rner}, J., {et~al.} 2022, \bibinfo{title}{{Regimes of cosmic-ray diffusion in Galactic turbulence},} SN Applied Sciences, 4, 15, \dodoi{10.1007/s42452-021-04891-z}

\bibitem[{M. {Ruszkowski} \& C. {Pfrommer}(2023){Ruszkowski} \& {Pfrommer}}]{Ruszkowski2023}
{Ruszkowski}, M., \& {Pfrommer}, C. 2023, \bibinfo{title}{{Cosmic ray feedback in galaxies and galaxy clusters},} \aapr, 31, 4, \dodoi{10.1007/s00159-023-00149-2}

\bibitem[{M.~L. {Sampson} {et~al.}(2023){Sampson}, {Beattie}, {Krumholz}, {Crocker}, {Federrath}, \& {Seta}}]{Sampson2023}
{Sampson}, M.~L., {Beattie}, J.~R., {Krumholz}, M.~R., {et~al.} 2023, \bibinfo{title}{{Turbulent diffusion of streaming cosmic rays in compressible, partially ionized plasma},} \mnras, 519, 1503, \dodoi{10.1093/mnras/stac3207}

\bibitem[{A.~A. {Schekochihin} {et~al.}(2004){Schekochihin}, {Cowley}, {Taylor}, {Maron}, \& {McWilliams}}]{Schekochihin2004a}
{Schekochihin}, A.~A., {Cowley}, S.~C., {Taylor}, S.~F., {Maron}, J.~L., \& {McWilliams}, J.~C. 2004, \bibinfo{title}{{Simulations of the Small-Scale Turbulent Dynamo},} \apj, 612, 276, \dodoi{10.1086/422547}

\bibitem[{T. {Schorlepp} {et~al.}(2025){Schorlepp}, {Kormann}, {L{\"u}bke}, {Sch{\"a}fer}, \& {Grauer}}]{Schorlepp2025a}
{Schorlepp}, T., {Kormann}, K., {L{\"u}bke}, J., {Sch{\"a}fer}, T., \& {Grauer}, R. 2025, \bibinfo{title}{{Synthetic turbulence via an instanton gas approximation},} \pre, 112, 055108, \dodoi{10.1103/4hh3-r1v6}

\bibitem[{Z.-S. {She} \& E. {Leveque}(1994){She} \& {Leveque}}]{She1994}
{She}, Z.-S., \& {Leveque}, E. 1994, \bibinfo{title}{{Universal scaling laws in fully developed turbulence},} \prl, 72, 336, \dodoi{10.1103/PhysRevLett.72.336}

\bibitem[{D. {Tharakkal} {et~al.}(2023){Tharakkal}, {Snodin}, {Sarson}, \& {Shukurov}}]{Tharakkal2023}
{Tharakkal}, D., {Snodin}, A.~P., {Sarson}, G.~R., \& {Shukurov}, A. 2023, \bibinfo{title}{{Cosmic rays and random magnetic traps},} \pre, 107, 065206, \dodoi{10.1103/PhysRevE.107.065206}

\bibitem[{M. Wilbert(2023)Wilbert}]{WilbertPhd2023}
Wilbert, M. 2023, \bibinfo{title}{Implementation and application of a pseudo-spectral MHD solver combined with an immersed boundary method to support the DRESDYN dynamo experiment,} PhD thesis, Ruhr-Universität Bochum

\bibitem[{M. {Wilbert} {et~al.}(2022){Wilbert}, {Giesecke}, \& {Grauer}}]{Wilbert2022}
{Wilbert}, M., {Giesecke}, A., \& {Grauer}, R. 2022, \bibinfo{title}{{Numerical investigation of the flow inside a precession-driven cylindrical cavity with additional baffles using an immersed boundary method},} Physics of Fluids, 34, 096607, \dodoi{10.1063/5.0110153}

\bibitem[{Y.-W. {Xiao} {et~al.}(2025){Xiao}, {Zhang}, \& {Xu}}]{Xiao2025}
{Xiao}, Y.-W., {Zhang}, J.-F., \& {Xu}, S. 2025, \bibinfo{title}{{Studying the diffusion mechanism of cosmic-ray particles},} \aap, 699, A317, \dodoi{10.1051/0004-6361/202453340}

\bibitem[{S. {Xu} \& H. {Yan}(2013){Xu} \& {Yan}}]{Xu2013}
{Xu}, S., \& {Yan}, H. 2013, \bibinfo{title}{{Cosmic-Ray Parallel and Perpendicular Transport in Turbulent Magnetic Fields},} \apj, 779, 140, \dodoi{10.1088/0004-637X/779/2/140}

\bibitem[{H. {Yan} \& A. {Lazarian}(2004){Yan} \& {Lazarian}}]{Yan2004}
{Yan}, H., \& {Lazarian}, A. 2004, \bibinfo{title}{{Cosmic-Ray Scattering and Streaming in Compressible Magnetohydrodynamic Turbulence},} \apj, 614, 757, \dodoi{10.1086/423733}

\bibitem[{H. {Yan} \& A. {Lazarian}(2008){Yan} \& {Lazarian}}]{Yan2008}
{Yan}, H., \& {Lazarian}, A. 2008, \bibinfo{title}{{Cosmic-Ray Propagation: Nonlinear Diffusion Parallel and Perpendicular to Mean Magnetic Field},} \apj, 673, 942, \dodoi{10.1086/524771}

\bibitem[{Y. {Yang} {et~al.}(2019){Yang}, {Wan}, {Matthaeus}, {Shi}, {Parashar}, {Lu}, \& {Chen}}]{Yang2019}
{Yang}, Y., {Wan}, M., {Matthaeus}, W.~H., {et~al.} 2019, \bibinfo{title}{{Role of magnetic field curvature in magnetohydrodynamic turbulence},} Physics of Plasmas, 26, 072306, \dodoi{10.1063/1.5099360}

\bibitem[{S.~L. {Young} {et~al.}(2008){Young}, {Denton}, {Anderson}, \& {Hudson}}]{Young2008}
{Young}, S.~L., {Denton}, R.~E., {Anderson}, B.~J., \& {Hudson}, M.~K. 2008, \bibinfo{title}{{Magnetic field line curvature induced pitch angle diffusion in the inner magnetosphere},} Journal of Geophysical Research (Space Physics), 113, A03210, \dodoi{10.1029/2006JA012133}

\bibitem[{V. {Zaburdaev} {et~al.}(2015){Zaburdaev}, {Denisov}, \& {Klafter}}]{Zaburdaev2015}
{Zaburdaev}, V., {Denisov}, S., \& {Klafter}, J. 2015, \bibinfo{title}{{L{\'e}vy walks},} Reviews of Modern Physics, 87, 483, \dodoi{10.1103/RevModPhys.87.483}

\bibitem[{C. {Zhang} \& S. {Xu}(2023){Zhang} \& {Xu}}]{Zhang2023a}
{Zhang}, C., \& {Xu}, S. 2023, \bibinfo{title}{{Numerical Testing of Mirror Diffusion of Cosmic Rays},} \apjl, 959, L8, \dodoi{10.3847/2041-8213/ad0fe5}

\bibitem[{M. {Zhou} {et~al.}(2020){Zhou}, {Loureiro}, \& {Uzdensky}}]{Zhou2020}
{Zhou}, M., {Loureiro}, N.~F., \& {Uzdensky}, D.~A. 2020, \bibinfo{title}{{Multi-scale dynamics of magnetic flux tubes and inverse magnetic energy transfer},} Journal of Plasma Physics, 86, 535860401, \dodoi{10.1017/S0022377820000641}

\bibitem[{G. {Zimbardo} \& S. {Perri}(2020){Zimbardo} \& {Perri}}]{Zimbardo2020}
{Zimbardo}, G., \& {Perri}, S. 2020, \bibinfo{title}{{Non-Markovian Pitch-angle Scattering as the Origin of Particle Superdiffusion Parallel to the Magnetic Field},} \apj, 903, 105, \dodoi{10.3847/1538-4357/abb951}

\bibitem[{R. Zwanzig(1973)Zwanzig}]{zwanzig:1973}
Zwanzig, R. 1973, \bibinfo{title}{Nonlinear generalized Langevin equations,} Journal of Statistical Physics, 9, 215, \dodoi{10.1007/BF01008729}

\bibitem[{E.~G. {Zweibel}(2013){Zweibel}}]{Zweibel2013}
{Zweibel}, E.~G. 2013, \bibinfo{title}{{The microphysics and macrophysics of cosmic rays},} Physics of Plasmas, 20, 055501, \dodoi{10.1063/1.4807033}

\end{thebibliography}

\end{document}